\documentclass[fleqn,usenatbib]{mnras}

\usepackage{newtxtext,newtxmath}
\usepackage[T1]{fontenc}

\DeclareRobustCommand{\VAN}[3]{#2}
\let\VANthebibliography\thebibliography
\def\thebibliography{\DeclareRobustCommand{\VAN}[3]{##3}\VANthebibliography}

\usepackage{graphicx}	% Including figure files
\usepackage{amsmath}	% Advanced maths commands
\usepackage{CJK}
\usepackage{gensymb}

\newcommand{\sgra}{SgrA$^\ast$}
\def\m87{M87$^\ast$}

\title[Misaligned accretion disks]{Supermassive Black Holes: modelling strongly and weakly magnetised misaligned accretion disks}

\author[J. S. Stanway et al.]{
Joshua S. Stanway,$^{1}$\thanks{E-mail: jsstanway@lancashire.ac.uk}
Cora Prather,$^{2}$
Derek Ward-Thompson,$^{1}$\thanks{E-mail: dward-thompson@lancashire.ac.uk}
Timothy J. Walton,$^{1}$
Brett Patterson,$^{1}$ \newauthor
and Hyerin Cho $^{2,3}$
\\
$^{1}$Jeremiah Horrocks Institute, University of Lancashire, Preston PR1 2HE, UK\\
$^{2}$Black Hole Initiative at Harvard University, 20 Garden Street, Cambridge, MA 02138, USA\\
$^{3}$Center for Astrophysics $\vert$ Harvard \& Smithsonian, 60 Garden Street, Cambridge, MA 02138, USA
}

\date{Accepted XXX. Received YYY; in original form ZZZ}

\pubyear{\the\year{}}

\begin{document}
\label{firstpage}
\pagerange{\pageref{firstpage}--\pageref{lastpage}}
\maketitle

\begin{abstract}
In this paper, we carry out a numerical study of misaligned accretion disks around spinning supermassive black holes. Here, we conduct a parameter survey covering a range of initial disk misalignment angles ($\mathcal{T}_\mathrm{init}=15\degree, 45\degree, 75\degree$) with either the Magnetically Arrested Disk (MAD) or Standard And Normal Evolution (SANE) magnetic field configurations, using the general-relativistic magnetohydrodynamic (GRMHD) code KHARMA. We find that models in the MAD state can align with the black hole up to $\sim10 \, r_g$, even in extremely misaligned models ($\mathcal{T}_\mathrm{init}=75^\circ$), which has not been seen before. Models without a dynamically important magnetic field remain misaligned up to the black hole, with a maximum disk tilt at $\sim10 \, r_g$, the peak's radial distance from the black hole increases with increasing disk misalignment. However, the maximum disk tilt does not have a linear relationship with the initial disk misalignment, and appears to have a maximum value of $\sim50^\circ$. We also show misaligned disk simulations produced in KHARMA are consistent with other GRMHD codes, for a variety of problems.
\end{abstract}


%%%%%%%%%%%%%%%%%%%%%%%%%%%%%%%%%%%%%%%%%%%%%%%%%%

%%%%%%%%%%%%%%%%% BODY OF PAPER %%%%%%%%%%%%%%%%%%

\section{Introduction}

Misalignment between the Supermassive Black Hole (SMBH) spin axis and their accretion disk represents the most general case of accretion, since infalling material is unaware of the SMBH spin axis. If the angular momentum of the accretion disk and the black hole are misaligned, Lense-Thirring precession acts on the accretion disk, resulting in a warped and twisted accretion disk. 

The Event Horizon Telescope (EHT) has produced resolved total intensity \citep{M87_paper1, SgrA_paper1, M87_2018_paper1} and polarized \citep{M87_paper7, SgrA_paper7} images of the SMBH \m87 and Sagittarius A* (\sgra), unlocking a new frontier in black hole astrophysics. These observations allow direct comparisons to complex numerical models of black hole accretion disks (e.g. \citealt{M87_paper5, M87_paper8, SgrA_paper5, SgrA_paper8}). 

Generally, numerical models used to study \m87 and \sgra are initialised such that the angular momentum of the accretion disk and black hole spin are aligned. In cell-based General Relativistic Magnetohydrodynamics (GRMHD) this is advantageous because resolution can be concentrated towards the simulation's midplane, and away from the polar regions, significantly increasing the time step. However, infalling material is unaware of the black hole spin, so the accretion flow is often expected to be misaligned with respect to the black hole spin axis. A misaligned, or tilted accretion disk evolves in a significantly different way to an equivalent aligned disk model, as first shown in General Relativistic Hydrodynamic simulations \cite{Fragile_Anninos_2005}. 

Low luminosity Active Galactic Nuclei (AGN) such as \m87 and \sgra\footnote{\sgra is believed to be an AGN in a quiescent state \citep{Mou_2014}} are expected to have geometrically thick, Radially Inefficient Accretion Flows (RIAF, see \citealt{Yuan2014} for a review) due to their low mass accretion rate ($\dot{M}$). In the thick disk regime, the accretion disk is expected to remain misaligned at larger radii \citep{Fragile_2007, Liska_2018}. The \cite{Bardeen_1975} effect can align the disk with the black hole spin on timescales of $M / \dot{M}$, where $M$ is the black hole mass, which is much longer for geometrically thin disks \citep{White_2020}. For \sgra, this is approximately $10^{13} - 10^{15} \, \mathrm{yr}$ in its current quiescent state\footnote{$M \sim 10^6 \, M_\odot$ \citep{GRAVITY_2023} and $\dot{M} \sim 10^{-9} - 10^{-7} \, M_\odot \, \mathrm{yr}^{-1}$ \citep{Bower_2003, Marrone_2007}.}. In addition, a geometrically thick accretion disk will warp and twist due to \cite{Lense_Thirring_1918} (LT) precession. This has been confirmed in many numerical studies (see e.g. \citealt{Fragile_2007, Liska_2018, White_2019}). Magnetic torque from the jet can also act to precess the accretion disk \citep{Jiang_2025}.

%Many systems, including AGN \citep{Caproni_2006, Caproni_2007}, X-ray binaries \citep{Greene_2001, Maccarone_2002}, and precessing relativistic jets \citep{Cui_2023, Dominik_2021, Foschi_2025} provide observational evidence for misaligned accretion disks. We also expect observational signatures of misalignment to be present on horizon scale images of SMBHs. \cite{Chatterjee_2020} found Doppler boosting in plunging streams restricts the peak brightness location more than in aligned models, that are dominated by gravitational lensing with no present plunging streams. In misaligned models, the ring diameter at 230 GHz can be larger, and more variable \citep{White_2022}, while the ring itself more elliptical \citep{White_2020}.

We also expect observational signatures of misalignment to be present on horizon scale images of SMBHs. In misaligned models, the ring diameter at 230 GHz can be larger, and more variable \citep{White_2022}, while the ring itself more elliptical \citep{White_2020}. A full library comparison (e.g. \citealt{SgrA_paper5}), however, has not been completed for misaligned disk models due to the two additional degrees of freedom\footnote{Initial misalignment angle and azimuthal viewing angle. When the accretion disk is misaligned, azimuthal symmetry is lost.}, but image studies on smaller image libraries have found interesting results. Misaligned models scaled to the same mass density as \m87 have been found to match 2017 observations well \citep{Chatterjee_2020}. They have also been noted as a potential cause for the precessing jet launched by \m87 \citep{Cui_2023}. Although a limited sample of misaligned models were included by \cite{SgrA_paper5}, no strong constraints on the tilt could be made due to short integration time.  
GRMHD simulations of RIAF models typically initialise the magnetic field by one of two methods, the strong and dynamically important magnetically arrested disk (MAD; \citealt{Igumenshchev_2003, Narayan_2003}), as well as the weaker Standard And Normal Evolution (SANE; \citealt{Narayan_2012, Sadowski_2013}). Misaligned models typically use MAD \citep{Ressler_2020, Ressler_2023, Chatterjee_2023} or strong near-MAD magnetic fields \citep{Liska_2018, Chatterjee_2020}. Misaligned models with the SANE magnetic field configuration have not been investigated for extreme initial misalignment angles ($>60^\circ$, e.g. \citealt{Liska_2018,White_2020, White_2022, Jiang_2025}).

In this work, we present the first highly misaligned ($>60^\circ$) SANE simulations and investigate the difference between equally misaligned MAD simulations produced by different GRMHD codes. In Section \ref{sec::method}, we provide a summary of our numerical methods. We present our results in Section \ref{sec::results}, discuss our findings in Section~\ref{sec::discussion} and conclude our results in Section \ref{sec::conclusion}.

\section{Numerical Methods}
\label{sec::method}

The simulations in this study are produced with KHARMA\footnote{https://github.com/AFD-Illinois/kharma} (Kokkos-based High Accuracy Relativistic Magnetohydrodynamics with Adaptive Mesh Refinement) \citep{Prather_2024}, a GPU accelerated implementation of the High Accuracy Relativistic Magnetohydrodynamics (HARM) numerical scheme \citep{Gammie_2003}. We adopt gravitational units in which $G = c = 1$, the gravitational radius, $r_g = GM/c^2$, where $M$ is the black hole mass. 

Our simulations have a base resolution of $N_r \times N_\theta \times N_\varphi = 384 \times 192 \times 192$ and use Widepole (WKS) coordinates to increase the timestep. In WKS coordinates we take logarithmic spacing in the radial direction, $x^r=\log r$, equal spacing in the azimuthal direction, $x^\varphi = \varphi$, and the theta grid is modified such that, $x^\theta = \theta$, where
\begin{multline}
    \theta = \frac{\pi}{2} [ 1 + f_\mathrm{lin} (2 x^\theta - 1) + (1 - f_\mathrm{lin}) \\
    \times \left\{ \tanh{\left( \frac{x^\theta - 1}{\lambda} \right)} + 1 \right\} - (1 - f_\mathrm{lin}) 
    \left\{ \tanh{ \left( - \frac{x^\theta}{\lambda} \right)} + 1 \right\} \biggr],   
\end{multline}
we set $f_\mathrm{lin} = 0.6$ and $\lambda = 0.03$, $x^r, x^\theta, x^\varphi$ are the evenly spaced code coordinates. A detailed discussion on WKS coordinates can be found in \citealt[Appendix C]{Cho_2024}. The radial grid is constructed such that 5 zones are placed within the EH, regardless of resolution, and we set the outer edge of the simulation domain to $r=1000r_g$. The theta and phi grids cover the ranges $[0, \pi]$ and, $[0, 2\pi]$ respectively. The grid is derefined by a factor of 4 at the pole using Internal Static Mesh Refinement (ISMR) with the implementation described in \cite{Cho_2025}.

We initialise our simulations with a hydrostatic Fishbone-Moncrief torus (FM; \citealt{FM_1976}), initially misaligned by either $15^\circ$, $45^\circ$, $75^\circ$ around a Kerr black hole with $a_\ast=0.9375$. We implement \texttt{H-AMR} style polar boundary conditions (\citealt[Supplementary Information 2.3]{Liska_2018}) to enable large misalignment angles. We describe and test our boundary implementation in Appendix~\ref{sec::appendix}. Each simulation is evolved to $30,000 t_g$, where $t_g = r_g/c$. We assume an ideal gas equation of state, with an adiabatic index of $\gamma = 5/3$. The electromagnetic vector potential, $A_\phi$, is initialised for MAD models such that,
\begin{equation}
    A_\phi = \mathrm{max} \Bigr[ \frac{\rho}{\rho_\mathrm{max}} \Bigr(\frac{r}{r_0} \sin \theta \Bigr)^3 e^{-r/400} - 0.2, 0 \Bigr],
    \label{eq::MAD}
\end{equation}
and in SANE models,
\begin{equation}
    A_\phi = \mathrm{max} \Bigr[ \frac{\rho}{\rho_\mathrm{max}} - 0.2, 0 \Bigr],
    \label{eq::SANE}
\end{equation}
where $\rho$ is the gas density, $\rho_\mathrm{max}$ is the maximum initial plasma density and $r_0$ is the inner edge of the accretion disk (\citealt{Wong_2022} and references therein). We normalise the magnetic pressure such that $\beta_\mathrm{min} \equiv P_\mathrm{gas} / P_\mathrm{mag} = 100$. We place the inner edge of the accretion disk at $20 r_g$ ($10  r_g$) and the pressure maximum at $41 r_g$ ($20 r_g$) for MAD (SANE) simulations.

\begin{figure*}
    \centering
    \includegraphics[width=\linewidth]{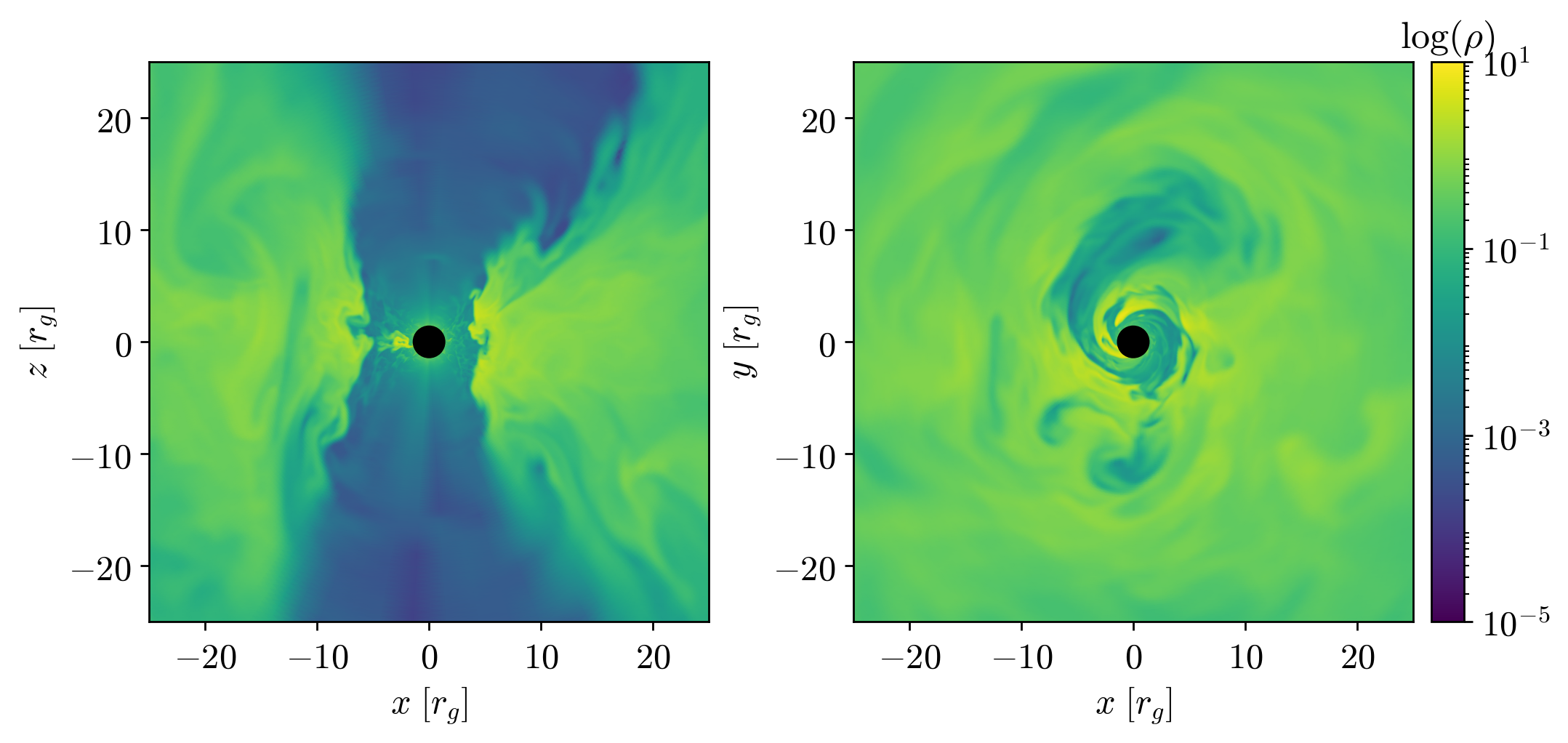}
    \caption{Snapshot images during a flux eruption event during the late time evolution of our \texttt{MAD45} model. The left panel plots a vertical cross-section, and the right a horizontal cross-section. The colour map indicates logarithmic gas density over six orders of magnitude.}
    \label{fig:snapshot}
\end{figure*}

To ensure stability we apply the following limits on the fluid: density $\rho > 10^{-5} \, r^{-3/2}$, internal energy density $u > 10^{-7} \, r^{-3/2}$, magnetisation $\sigma < 10^2$, gas to magnetic pressure ratio $\beta^{-1} < 10^3$, and the maximum Lorentz factor $\Gamma < 10$. KHARMA uses a first-order flux correction scheme \citep{Beckwith_2011} and a Kastaun inverter \citep{Kastaun_2021} to allow for more lenient floors, and improved stability. Fluxes are calculated with the Local Lax-Friedrichs (LLF) method \citep{Friedrichs_1971}, and the Face-CT scheme \citep{Gardiner_2008} is used to enforce no divergence of the magnetic field. Flux reconstruction uses a linearised version of the WENO-Z \citep{Castro_2011} scheme. The allowed timestep is set by the Courant-Friedrichs-Lewy (CFL) condition \citep{cfl_1967}, which we set as CFL=0.8. Throughout this work, each model follows the naming convention $\mathrm{[Magnetic \ Field \ Configurtion]  [\mathrm{Inital \ Misalignment \ Angle}]}$, e.g. the model with the MAD magnetic field configuration, initial misaligned by 45$^\circ$ is named \texttt{MAD45}. Figure~\ref{fig:snapshot} shows a snapshot of \texttt{MAD45} during a flux eruption event at late times.

\section{Results}
\label{sec::results}

In this work, we present results from six different simulations, varying the initial misalignment angle of the accretion disk, and magnetic field structure. We first discuss the effects on the accretion disk, and follow this with a discussion of the jet in our misaligned models. A summary of time averaged results are presented in Table~\ref{tab:Time-averaged}. 

\begin{table}
    \centering
    \caption{Time averaged quantities over the final $5,000 t_g$ of each simulation. The quantities are as follows: the mass accretion rate, $\dot{M}$; dimensionless magnetic flux, $\Phi_\mathrm{BH} / \sqrt{\langle \dot{M} \rangle}$; outflow efficiency, $\eta_\mathrm{out}$; barycentric disk radius, $R_\mathrm{disk}$; disk tilt angle, $\mathcal{T}_\mathrm{disk}$, measured at $10r_g$ and $200r_g$; and jet tilt angle, $\mathcal{T}_\mathrm{jet}$, measured at $10r_g$, and $200r_g$. For readability, we shorten the magnetic field configuration to either ``M'' or ``S''.}
    \renewcommand{\arraystretch}{1.25}
    \begin{tabular}{c|ccc}
        \hline
        \hline
        Time averaged &  & Model &  \\
        quantity & $\mathrm{M15/S15}$ & $\mathrm{M45/S45}$ & $\mathrm{M75/S75}$ \\
        \hline
        $\langle \dot{M} \rangle$ & 24.5 / 1.5 & 15.8 / 2.9 & 15.4 / 1.2  \\
        $\langle \Phi_\mathrm{BH} / \sqrt{ \dot{M}} \, \rangle$ & 17.3 / 1.3 & 16.1 / 1.3 & 17.3 / 1.1 \\
        $\langle \eta_\mathrm{out} \rangle \, [\%]$ & 208 / 3.3 & 179 / 2.1 & 207 / 1.6 \\
        $\langle R_\mathrm{disk} \rangle \, [r_g]$ & 535 / 135 & 550 / 120 & 555 / 133 \\
        $\langle \mathcal{T}_{\mathrm{disk}} \ (10) \rangle [^\circ]$ & 4.6 / 16.4 & 5.8 / 41.9 & 5.7 / 46.5 \\
        $\langle \mathcal{T}_{\mathrm{disk}}\ (200 /100) \rangle [^\circ]$ & 21.2 / 11.0 & 53.4 / 33.2 & 54.0 / 34.8 \\
        $\langle \mathcal{T}_{\mathrm{jet}} \ (10) \rangle [^\circ]$ & 6.9 / 19.4 & 8.1 / 42.6 & 8.0 / 41.1 \\
        $\langle \mathcal{T}_{\mathrm{jet}}\ (200 / 100) \rangle [^\circ]$ & 8.8 / 15.7 & 15.8 / 34.0 & 11.4 / 42.8 \\
        \hline
    \end{tabular}
    \label{tab:Time-averaged}
\end{table}

\subsection{Misaligned Magnetically Arrested Disks}

\begin{figure*}
    \centering
    \includegraphics[width=\linewidth]{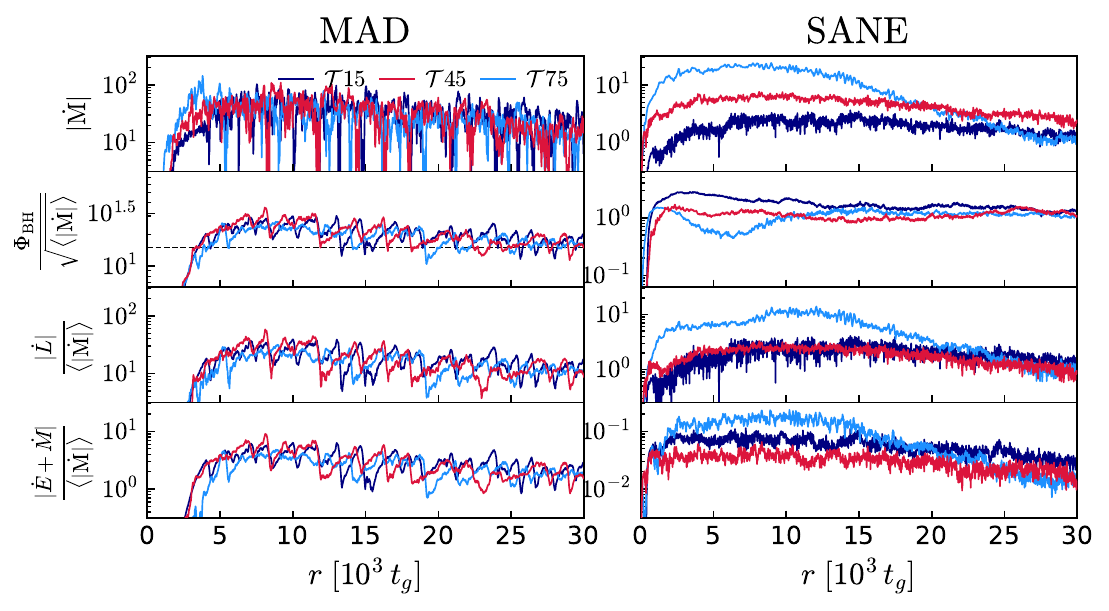}
    \caption{Time series of radial fluxes for our six simulations. From top to bottom, rest mass accretion rate, normalised dimensionless horizon penetrating magnetic flux, normalised angular momentum, and normalised energy outflow with rest mass contributions removed. MAD simulations are plotted in the left column, and SANE simulations in the right panel. Simulations initially misaligned by $15^\circ$ are plotted in navy, $45^\circ$ in crimson, and $75^\circ$ in light blue. The dimensionless magnetic flux for a model to develop into the MAD state, $\phi \gtrsim 15$, is shown by the dashed black line. MAD models become magnetically saturated at a much larger value of $\Phi_\mathrm{BH} / \sqrt{\langle | \dot{M} | \rangle}$ compared to SANE models, leading to episodic flux eruption events when excess magnetic flux is trapped on the event horizon. Flux eruption events are identified by spiking magnetic flux, angular momentum, and energy outflow, and a sharp decrease in angular momentum. SANE models do not form flux eruption events, so the evolution of radial fluxes is comparatively quite.}
    \label{fig:radial-fluxes}
\end{figure*}

We first consider the time dependent radial fluxes at the event horizon for our three MAD simulations. In Figure~\ref{fig:radial-fluxes} we plot the rest mass accretion rate $\dot{M} = \int_\phi \int_\theta - \rho u^r \sqrt{-g} \, d\theta \, d\phi$, dimensionless magnetic flux, $\phi = \Phi / \sqrt{\dot{M}}$, where $\Phi_{\mathrm{BH}} = \frac{1}{2} \int_\phi \int_\theta |B^r| \sqrt{-g} \, d\theta \, d\phi$, angular momentum $\dot{L} = \int_\phi \int_\theta T^r_\phi \sqrt{-g} \, d\theta \, d\phi$ and energy $\dot{E} = \int_\phi \int_\theta - T^r_t \sqrt{-g} \, d\theta \, d\phi$ fluxes, where $T^\mu_\nu$ is the stress energy tensor. $\Phi_{BH}$ and $\dot{L}$ are normalised by the accretion rate. We remove the rest mass contribution from the energy ($|\dot{E} + \dot{M}|$), and normalise by the average accretion rate over the final $5,000t_g$, this is equivalent to the outflow efficiency $\eta_\mathrm{out}$. 

To avoid contamination by density floors, $\dot{M}$, $\dot{E}$, and $\dot{L}$ are calculated at $r=5r_g$. $\phi_\mathrm{BH}$ is unaffected by density floors and is therefore calculated at the event horizon $r_\mathrm{EH} = r_g(1 + \sqrt{1 - a_\ast^2})$. For each model, we calculate the average barycentric disk radius, $R_\mathrm{disk}$, (Table~\ref{tab:Time-averaged}; see also \citealt{Porth_2019}), radial profiles throughout Section~\ref{sec::results} are calculated over the radial domain $r \in [r_\mathrm{EH}, 500 r_g]$ for MAD models and $r \in [r_\mathrm{EH}, 100 r_g]$ for SANE models. This ensures the averages are only calculated over the disk.

Figure~\ref{fig:radial-fluxes}, despite large initial misalignment differences, all of our MAD simulations reach magnetic flux saturation, $\langle \phi_\mathrm{BH} \rangle \sim 15$ (in Lorentz-Heaviside units), and develop into the MAD state. We see clear dips in the accretion rate, corresponding with a spike in the magnetic flux, outward angular momentum, and energy outflow, reminiscent of geometrically thick aligned simulations in the MAD state (e.g. \citealt{Narayan_2012, Dhruv_2025}). Each MAD simulation also contains highly efficient outflows, $\eta_\mathrm{out} > 1$, (see Table~\ref{tab:Time-averaged}) suggesting the presence of a powerful \cite{BZ_1977} jet, extracting angular momentum from the black hole, as expected for an accretion disk in the MAD state around a rapidly spinning black hole. 

\begin{figure*}
    \centering
    \includegraphics[width=\linewidth]{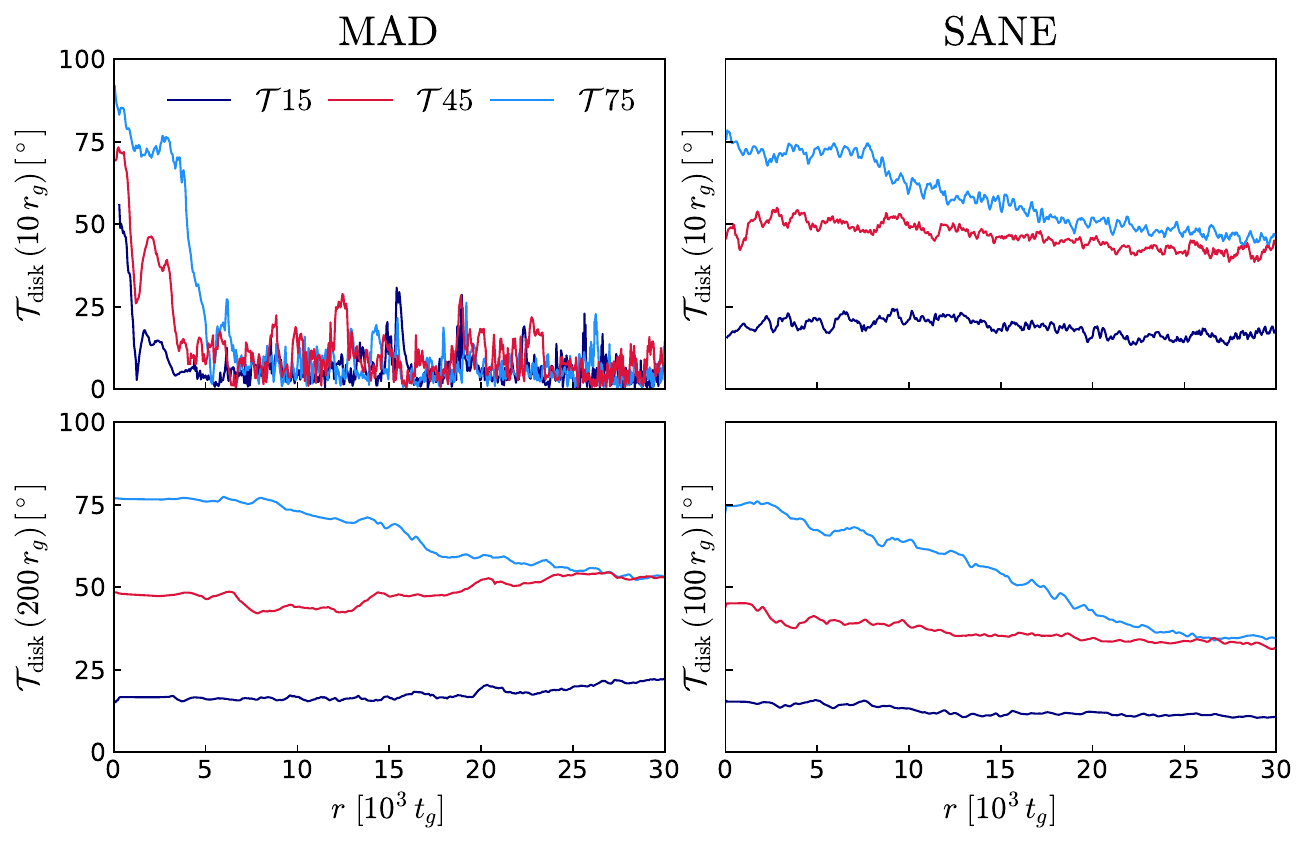}
    \caption{The powerful and efficient jet in MAD models quickly torques the inner edge of the accretion disk into alignment (top left) via the magneto-spin alignment mechanism, flux eruption events can, however, cause short-lived disk misalignment up to $\sim25\degree$. Without sufficient jet power, our SANE models cannot align with the black hole, although, we find the peak disk tilt ($\sim10r_g$, top right) in our extremely misaligned model tends towards $\sim50\degree$. The bottom row shows, at large radii, outside the presence of the jet, MAD and SANE models act similarly. Simulations initially misaligned by $15\degree$ are plotted in navy, $45\degree$ in crimson, and $75\degree$ in light blue.}
    \label{fig:disk-temporal-profile}
\end{figure*}

Figure~\ref{fig:disk-temporal-profile} plots the temporal evolution of the disk misalignment in Figure~\ref{fig:disk-temporal-profile} at $10 r_g$ (top left) and $200 r_g$ (bottom left) for our three MAD models. We measure the disk misalignment angle using the prescription described by \cite{Fragile_Anninos_2005} and \cite{Fragile_2007}. Regardless of the initial disk misalignment, the inner edge of each MAD accretion disk is torqued into alignment via magneto-spin alignment within the first $\sim5,000t_g$ of each simulation. Magnetic flux eruption events can, however, cause short-lived misalignment up to $\sim25^\circ$. At larger radii, without the jet torque acting on the accretion disk, the disk evolves over much longer timescales, with minimal or no alignment occurring during the simulation.

\begin{figure*}
    \centering
    \includegraphics[width=\linewidth]{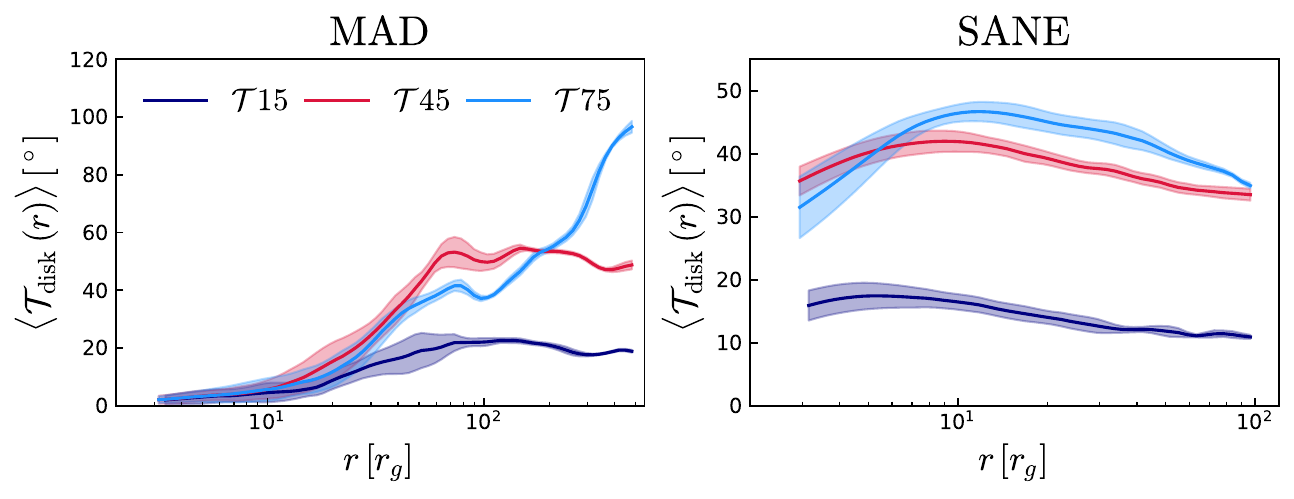}
    \caption{Radial profiles of the disk tilt in our MAD (left) and SANE (right) simulations, time averaged over the final $5,000 t_g$, with $1 \, \sigma$ error bars. Simulations initially misaligned by $15^\circ$ are plotted in navy, $45^\circ$ in crimson, and $75^\circ$ in light blue. All our MAD models align up to $r \gtrsim 10 r_g$ via the Magneto-spin alignment process, before increasing towards the initial misalignment angle. SANE models without a powerful jet do not align via the Magneto-spin alignment, but show a peak at approximately $10 r_g$, the position of this peak increases with increasing initial misalignment.}
    \label{fig:disk-radial-profile}
\end{figure*}

Figure~\ref{fig:disk-radial-profile} plots the time averaged radial profiles of our three MAD models in the left panel. In each MAD models, regardless of initial inclination, the accretion disk aligns with the black hole up to $r \gtrsim 10 r_g$. This is particularly interesting for the \texttt{MAD75} model, as previous work on extremely misaligned MAD models did not find near horizon alignment for large initial disk tilts $\mathcal{T} \geq 75^\circ$ \citep{Chatterjee_2023}. We further explore this model and the differences between extremely misaligned models in \cite{Chatterjee_2023} in Section~\ref{sec::extreme-misalinged}.

\subsection{Maximum Stable Disk Tilt around a rapidly rotating Kerr black hole}

SANE models have weak and inefficient jets, $\eta_\mathrm{out} \sim 10^{-2}$, much like aligned counterparts \citep{Narayan_2012, Dhruv_2025}. Weak jets do not supply sufficient torque on the disk to force the inner edge into alignment, and only LT precession acts to align the accretion disk, unlike MAD models, where the jet supplies sufficient torque to align the accretion disk.

In Figure~\ref{fig:disk-temporal-profile}, the right panels plot the evolution of disk misalignment throughout each simulation, close to the black hole, $10 r_g$, (top right) and at larger radii, $100r_g$, (bottom right). We observe minimal alignment for low and intermediate misaligned disks near the black hole, and our extremely misaligned SANE model slowly tends towards $\sim50^\circ$ near the black hole. At larger radii, the misalignment of our accretion disk decreases for all our SANE models, initially dramatically for the extremely misaligned SANE model before the angle stabilises at $\sim 25t_g$. The temporal evolution of disk misalignment at large radii from the black hole, beyond the effects of the jet, is similar for MAD and SANE misaligned disks for all initial misalignment angles in our simulations. 

Figure~\ref{fig:disk-radial-profile} (right panel) shows our SANE models do not align with the black hole at any radii, regardless of the initial disk misalignment. Within $10 r_g$, the accretion disk aligns marginally due to LT precession, with a peak misalignment angle at approximately $10 r_g$, as seen in other works \citep{Fragile_2007,Liska_2018}. The difference between the $\sim 10 r_g$ peak misalignment angle and, and the misalignment angle at the inner edge of the accretion increases with initial disk tilt. To such a degree, the inner edge of the \texttt{SANE75} accretion disk is more aligned than \texttt{SANE45}. Although, we note the larger $1\sigma$ uncertainty, so this will vary over the course of the simulation. Continuing radially from the $\sim10 r_g$ peak, the misalignment angle steadily decreases. These results are consistent with weak field models from previous studies \citep{Fragile_2007, Liska_2018, White_2019,Chatterjee_2020}, but we have now confirmed this to the largest initial misalignment angle in the existing literature. 

The \texttt{SANE75} model is perhaps not as misaligned as expected, the radial profile in Figure~\ref{fig:disk-radial-profile} is close to that of the \texttt{SANE45} model. It is not simply the difference in initial disk misalignment, as the \texttt{SANE15} model has a significantly different radial profile to the \texttt{SANE45} model, with the same difference in initial disk misalignment. In misaligned accretion disks with a weak or SANE magnetic field configuration, there appears a peak stable misalignment angle of $\sim50^\circ$. This argument is supported by the \texttt{T60} model from \cite{Chatterjee_2020}, initially misaligned by $60^\circ$, where the peak misalignment angle of the accretion disk is also $\sim50^\circ$ at $\sim10 r_g$.

To increase confidence in our results, we calculate MRI $\mathcal{Q}$-factors following the prescription in \cite{Porth_2019}. $\mathcal{Q}$-factors for our SANE models are presented in Table~\ref{tab:Qfactors}. The value of $\mathcal{Q}$-factors in each dimension suggest we have appropriate resolution for our problem size, and the MRI turbulence is captured accurately (\citealt{Hawley_2011, Hawley_2013, Porth_2019}, and references therein).  

\begin{table}
    \centering
    \caption{Time averaged MRI quality factors, $\langle \mathcal{Q}^{(r, \theta, \phi)} \rangle$, within $150 r_g$ of the black hole, over the final $5,000 t_g$ of each SANE simulation. $\mathcal{Q}$-factors quantify the number of cell available to resolve the fastest growing MRI mode in each dimension. We find sufficient resolution for our problem size (see \citealt{Porth_2019}, and reference therein). In the left column we have each SANE model, when the number represents the initial disk misalignment.}
    \begin{tabular}{c|c}
    \hline\hline
    Model & $(\langle \mathcal{Q}^{(r)} \rangle, \langle \mathcal{Q}^{(\theta)} \rangle, \langle \mathcal{Q}^{(\phi)} \rangle)$  \\
    \hline
    \texttt{SANE15}  & (15, 20, 21) \\
    \texttt{SANE45}  & (17, 34, 23) \\
    \texttt{SANE75}  & (15, 30, 17) \\
    \hline
    \end{tabular}
    \label{tab:Qfactors}
\end{table}

Models with a smaller initial disk misalignment (e.g. \texttt{SANE15}, \texttt{SANE45}, see also models in \citealt{Liska_2018, White_2019, Chatterjee_2020}) have a $\sim10 r_g$ peak at approximately equal, or sometimes slightly more than the initial misalignment angle. The peak misalignment and twisting of the inner accretion disk in SANE models is an expected cause of LT precession predicted in \cite{Lubow_2000, Lubow_2002}. The strength of LT precession is proportional to the spin of the black hole \citep{Lense_Thirring_1918, Bardeen_1975}, and therefore, accretion disks in the SANE state around a slower spinning black hole may have a higher peak misalignment angle. Both the $\sim 50^\circ$ peak misalignment and spin dependence of this peak angle could be confirmed with additional SANE simulations at higher initial misalignment angles, and simulations with slower spinning black holes. However, this is beyond the scope of this work.

We also observe this behaviour in our highly misaligned MAD model, \texttt{MAD75}, in the bottom left panel of Figure~\ref{fig:disk-temporal-profile}. At larger radii, the jet's influence on this disk is minimal, suggesting this effect is caused by the rapidly rotating black hole, and LT precession. In our upcoming work, we study how the magnitude and sign of the black hole spin impacts the evolution of misaligned accretion disks in the MAD state. In our slower spinning, and retrograde models, we do not observe the same effects as seen in \texttt{MAD75}, suggesting the maximum stable disk misalignment angle depends on the black hole spin. Slower spinning, and retrograde models, initially misaligned by $75\degree$, remain within $10\%$ of the initial disk tilt, over the duration of the simulation \citep{Stanway_in_prep}, compared to decreasing by $\sim33\%$ in \texttt{MAD75}.

\subsection{Tilted Jets in Misaligned Simulations}

\begin{figure*}
    \centering
    \includegraphics[width=\linewidth]{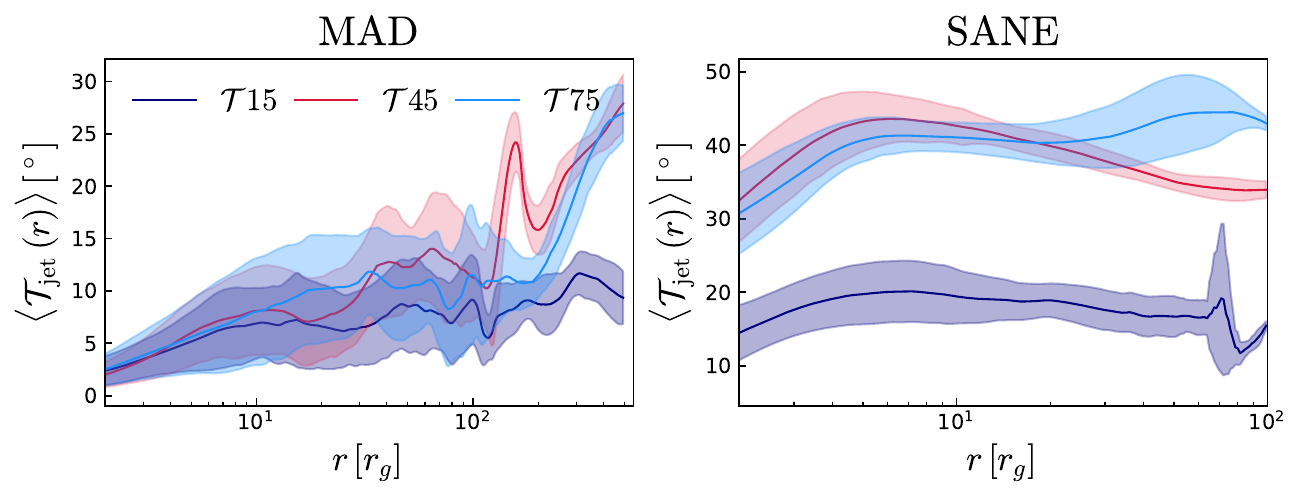}
    \caption{Radial profiles of the jet tilt in our MAD (left) and SANE (right) simulations, time averaged over the final $5,000 t_g$, with $1 \, \sigma$ error bars. Simulations initially misaligned by $15^\circ$ are plotted in navy, $45^\circ$ in crimson, and $75^\circ$ in light blue. In general, the jet follows the disk, in MAD models where the disk is aligned near the black hole, the jet has near zero tilt, increasing with radius as the disk tilt also increases. The accretion disk in SANE models never fully align with the black hole, which we also find to be true for the jet.}
    \label{fig:jet-radial}
\end{figure*}

Each of our three MAD simulations produce very powerful and efficient jets ($\eta_\mathrm{out} > 1$, see Figure~\ref{fig:radial-fluxes}, Table~\ref{tab:Time-averaged}). We measure the jet's position in each of our simulations following the \cite{Liska_2018} prescription, isolating the jet region with the criteria, $r p_\mathrm{mag} / \rho > 0.5$, where $p_\mathrm{mag}$ is the magnetic pressure. We then calculate the jet centre by calculating the average $x, y, z$ coordinates weighted by the magnetic pressure. The jet is separated into the lower and upper jet, for our analysis we only consider the upper jet. The time-averaged radial profile of the jet in our simulations is shown in Figure~\ref{fig:jet-radial}.

Figure~\ref{fig:jet-radial} shows jet alignment near the black hole in each MAD model. This is expected with alignment of the inner edge of the accretion disk, as the jet is launched along the spin axis of the black hole where no material is present. However, as the radial distance increases, the accretion disk is no longer in alignment, causing significantly more jet-disk collisions compared to aligned simulations. This causes the jet to becomes misaligned with respect to the black hole spin vector at large radii. Our results agree with findings in similar works, e.g. \citealt{Liska_2018, Chatterjee_2020, Jiang_2025}.

Each MAD model, regardless of initial misalignment angle, follows approximately the same radial profile until $\sim100 r_g$, the jet's misalignment steadily increases towards $\sim10^\circ$. As the radial distance increases beyond this point, the jet's misalignment in \texttt{MAD45} and \texttt{MAD75} models rapidly increases, whereas the \texttt{MAD15} model follows the previous trend. This is expected given the radial disk profiles of each MAD model. We note at small radii the $1\sigma$ uncertainty is large, indicating significant precession in the jets for all models, we discuss prospects of observing jet precession in future EHT movie campaigns in Section~\ref{sec::jet-precession}

Our SANE models, on the other hand, produce very weak jets, with efficiencies on the order of $1\%$ (see Table~\ref{tab:Time-averaged}), consistent with weakly magnetised aligned models \citep{Narayan_2012, Dhruv_2025}. Much like the jet in MAD models, the jet in SANE models closely follows the radial disk profile in Figure~\ref{fig:disk-radial-profile}. In SANE models, however, the jet never aligns with the black hole, remaining misaligned up to the black hole. Our results agree with previous findings \citep{Liska_2018, Chatterjee_2020}. 

Our SANE models show decreasing jet efficiency as the initial misalignment increases (see Table~\ref{tab:Time-averaged}), the outflow in \texttt{SANE15} is approximately twice as efficient as in \texttt{SANE75}. This could have an interesting impact on black hole spin evolution. \cite{Gammie_2004} calculated the spin equilibrium for weakly magnetised SANE black holes to be $a_\ast \approx 0.93$. The SANE models in this work suggest there may be a strong dependence on misalignment for the spin equilibrium for weakly magnetised black holes. Unfortunately, we cannot calculate the spin equilibrium, as our weakly magnetised misaligned models have only been produced with a single black hole spin. We see no clear link between the jets efficiency and disk misalignment for our MAD models, but that is likely due to the short averaging period. For example, the average jet efficiency in \texttt{MAD45} is the lowest across our MAD models, but this model has the fewest magnetic flux eruption events which cause a significant spike in the efficiency. A future misaligned library with longer duration simulations, and a denser spin coverage would undoubtedly discover interesting properties of the spin evolution in black holes, but this is beyond the scope of this work.

\section{Discussion}
\label{sec::discussion}

\begin{figure*}
    \centering
    \includegraphics[width=\linewidth]{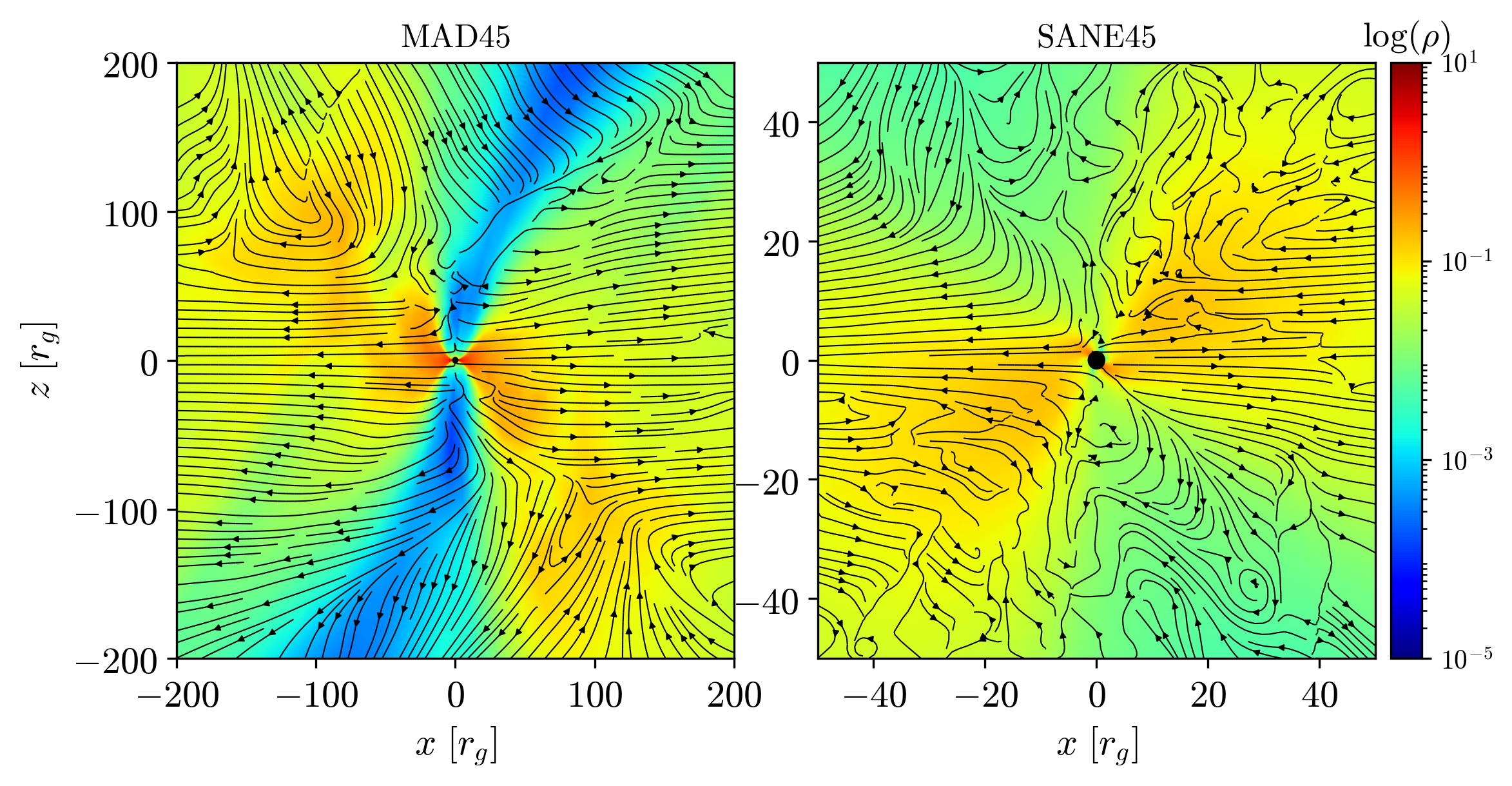}
    \caption{Ordered and chaotic magnetic field lines in our MAD and SANE simulations. The time-averaged logarithmic density of our \texttt{MAD45} (left) and \texttt{SANE45} (right) models in the $x-z$ plane. The overlaid streamlines are the time-averaged poloidal magnetic field lines. Time averaging is calculated over the range $t=25,000M$ to $t=30,000M$.}
    \label{fig:field-lines}
\end{figure*}

In Figure~\ref{fig:field-lines}, we plot the time-averaged poloidal magnetic field lines over the logarithmic density for the models \texttt{MAD45} (left panel) and \texttt{SANE45} (right panel). The strength and order of the magnetic field are responsible for the key differences between MAD and SANE models, much like in aligned counterparts. In each model, the rapidly spinning black hole can force the initial misaligned magnetic field into the equatorial plane via frame-dragging. 

Beyond the equatorial plane, the poloidal magnetic field within \texttt{MAD45} remains far more ordered than in \texttt{SANE45}, resulting in the formation of the powerful jet \citep{Narayan_2012}. This, in turn, results in magneto-spin alignment aligning the inner edge of the accretion disk \citep{Chatterjee_2023}. The disordered field in \texttt{SANE45} cannot form a powerful jet (see Table~\ref{tab:Time-averaged}, thus the disk remains misaligned. 

\subsection{The Extremely Misaligned Case}
\label{sec::extreme-misalinged}

Our findings throughout this work agree with the existing literature in all but one case, our extremely misaligned MAD model, \texttt{MAD75}. \citep{Chatterjee_2023} found models misaligned by such an extreme initial misalignment cannot reach the MAD state, forcing the inner accretion disk into alignment. We, however, found that it is possible to force the inner disk into alignment for the same extreme initial conditions. Here, we discuss potential causes for difference evolution between models.

First, we will consider the similarities between our model, \texttt{MAD75}, and \texttt{T75}. Both models initialize an FM torus with the same inner edge ($r_\mathrm{in}=20$), pressure maximum ($r_\mathrm{max}=41$), and an identical MAD magnetic field configuration (Eq.\ref{eq::MAD}) around a Kerr black hole with $a_\ast=0.9375$. Each model normalises the magnetic field strength by setting the ratio of maximum gas and magnetic pressure to 100. 

Our simulations differ in four key ways: the GRMHD code used, simulation resolution, duration, and the adiabatic index in the accretion disk. KHARMA's predecessor \texttt{iharm3d} \citep{Prather_2021}, and \texttt{H-AMR} both participated in detailed GRMHD code comparison projects \citep{Porth_2019, Olivares_MAD}, which found minimal differences between codes with identical (or as near as could be) setups. Misaligned disk simulations add complexity compared to standard aligned simulations, and no comprehensive code comparison exists, however, we do not expect different codes to be solely responsible for the differences we see between \texttt{MAD75} and \texttt{T75}.

The simulation resolution and duration are higher and longer in \texttt{T75}. While \texttt{MAD75} is at a lower resolution, we see outwards angular momentum is present (Figure~\ref{fig:radial-fluxes}), suggesting sufficient resolution to resolve MRI. Additionally, in the MAD regime, time averaged profiles appear resolution invariant \citep{Salas_2024, Olivares_MAD}. We also find our SANE models have sufficient resolution (see Table~\ref{tab:Qfactors}), even with a smaller disk size. Therefore, we do not believe these are responsible for the observed differences. The longer duration may be a factor in the evolutionary differences. \texttt{T75} reached a weakly magnetised state at late times, but unlike our \texttt{MAD75} model, it was never in a consistent MAD state. Therefore, we do not expect \texttt{MAD75} to reach the weakly magnetised state, even if it was run for an equivalent duration. 

The final key difference between \texttt{MAD75} and \texttt{T75} is the adiabatic index. To reduce computational cost, most GRMHD simulations evolve the total internal energy, $u = u_e + u_i$, where $u_e$ and $u_i$ are the internal energy of electrons and ions respectively, and assume an ideal equation of state with a fixed adiabatic index $\gamma$. The most common values in literature are $\gamma = 4/3, 13/9, 5/3$, each value leads to significant evolutionary differences. Recent works have suggested $\gamma = 5/3$ is the best choice \citep{Chael_2025, Gammie_2025}. The simulations in this work use $\gamma = 5/3$, whereas, \texttt{T75} uses $\gamma=13/9$. \cite{Narayan_2022,Begue_2023} have shown models with $\gamma=5/3$ have higher dimensionless magnetic flux at the horizon, $\phi$, compared to models with $\gamma = 4/3, 13/9$. The higher magnetic flux crossing the horizon produces the MAD state in \texttt{MAD75} (Figure~\ref{fig:radial-fluxes}), which does not form in \texttt{T75} (see Figure 2 in \citealt{Chatterjee_2023}). The MAD state, and the powerful jet it produced in \texttt{MAD75}, forces the inner edge of the accretion disk into alignment as seen in Figure~\ref{fig:disk-temporal-profile} and \ref{fig:disk-radial-profile}. We explore the impact of varying initial conditions, including the adiabatic index, in Appendix~\ref{sec::AppendixB}. We find changing the adiabatic index can drastically alter the evolution of a misaligned accretion disk. Future simulations should be as realistic as possible when selecting the adiabatic index. 

Our general findings, however, agree with \cite{Chatterjee_2023}, and our less tilted MAD models, \texttt{MAD15} and \texttt{MAD45}, match closely to equivalent models by \cite{Chatterjee_2023}. Magneto-spin alignment is clearly a complex process, depending not only on the magnetic field state and spin of the black hole, but also the adiabatic index of the plasma. It is likely, if we increased our initial misalignment beyond 75$\degree$, the disk would not align, and would display similar behaviour to extremely misaligned models by \cite{Chatterjee_2023}. Further work is clearly required to determine the specifics of magneto-spin alignment. In particular radiative GRMHD simulations, evolving the electron and ion populations individually (e.g. \citealt{Ressler_2015, Chael_2025}), would shine an interesting light on the process; however, this is beyond the scope of this work. We are, however, unaware of any geometrically thick, highly misaligned, simulations produced utilising radiative GRMHD.

\subsection{Prospects of Observing Jet Misalignment}
\label{sec::jet-precession}

One of the most likely methods of identifying disk misalignment is observing a twisted and precessing jet. On large scales we have observed relativistic jets from SMBHs precessing \citep{Cui_2023, Dominik_2021, Foschi_2025}, but this may be caused by external factors. If, however, the jet is observed to precess near the black hole, a misaligned accretion disk is the most likely cause. With greatly improved intermediate baselines in 2021 EHT observations of \m87, the first hint of non ring emission was detected. This is thought to be the base of the jet connecting to the ring \citep{M87_2021, Saurabh_2025}. As the EHT array continues to improve, it is expected the base of the jet can be resolved up to $100 \mu \mathrm{as}$ from \m87. In future observing campaigns, the EHT collaboration hopes to observe \m87 over $2-3 $ months \citep{EHT_white_paper}, equivalent to approximately $25 r_g$ and $150 - 250  t_g$ for \m87. %If these goals are met, what are the chances of observing jet precession, and can the angle of disk tilt be determined from near horizon jet precession? 

\begin{figure*}
    \centering
    \includegraphics[width=\linewidth]{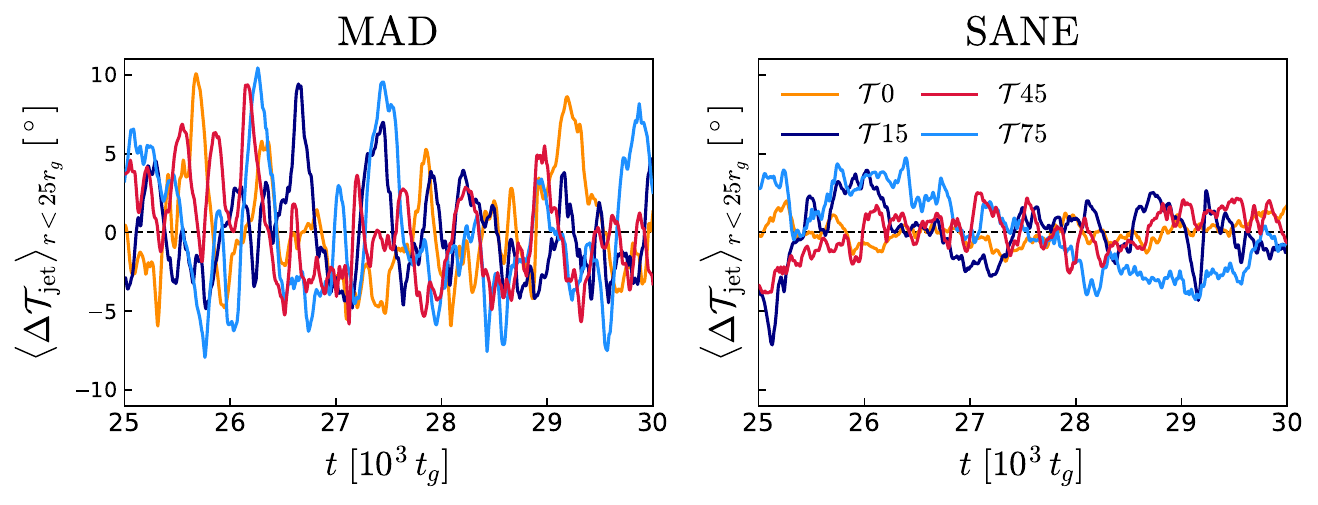}
    \caption{Temporal evolution of the mean subtracted jet tilt averaged within $25 r_g$ of the black hole. MAD models are shown in the left panel, and SANE models in the right panel. Simulations initially misaligned by $15^\circ$ are plotted in navy, $45^\circ$ in crimson, and $75^\circ$ in light blue, aligned models are shown in orange. MAD models show short, but drastic increases in jet tilt caused by magnetic flux eruption events, whereas, SANE models have much longer-term variations, on smaller scales.}
    \label{fig:jet-temporal}
\end{figure*}

First, consider how the jet evolves in our misaligned simulations. For this section, we also include two aligned simulations from Illinois v5 GRMHD library as a point of comparison \citep{Bowden_in_prep}. They are close matches to the misaligned disk simulations produced for this work, with the same resolution, disk size, magnetic field configuration, adiabatic index, and produced with KHARMA. We plot the mean subtracted temporal evolution of jet tilt averaged within $25 r_g$ of the black hole in Figure~\ref{fig:jet-temporal}.

In Figure~\ref{fig:jet-temporal} we plot the mean subtracted temporal evolution of the jet tilt, averaged within $25 r_g$, and find minimal differences between aligned and misaligned models. This is not unexpected for our MAD models. As shown in Figure~\ref{fig:disk-radial-profile} the inner edge of the accretion disk is aligned for all MAD models, disk misalignment, and therefore jet tilt does not significantly increase until $r \gtrsim 50 r_g$ (see also Figure~\ref{fig:jet-radial}). 

The jet in SANE models, on the other hand, have much less variation in the jet's angle, as we would expect. Magnetic flux eruption events are the main cause of the large variation in MAD models, which can alter the trajectory of the jet on short timescales. Without flux eruption events in SANE models, the jet evolves over much larger timescales, and smaller deviations by a factor of $\sim2$. The deviation from the mean angle in SANE models is generally within $5^\circ$ from the mean, whereas MAD models have regular deviations of $\sim10^\circ$ from the mean. A future \m87 movie campaign in which the jet base is recovered up to $25 r_g$ may not provide evidence of a misaligned accretion disk, but it may provide a powerful new constraint to differentiate MAD and SANE models.

\section{Conclusions}
\label{sec::conclusion}

In this work, our goal was to study the evolution of minimally, moderately, and extremely misaligned accretion disks in the MAD and SANE state. For this, we produced six 3D GRMHD simulations of accretion disks with the standard MAD (Eq.\ref{eq::MAD}) and SANE (Eq.\ref{eq::SANE}) magnetic field configurations, and initial disk misalignment angles of $\mathcal{T}_\mathrm{init}=15^\circ, 45^\circ, 75^\circ$, around a Kerr black hole with spin parameter $a_\ast=0.9375$. We summarise our results here: 
\begin{itemize}

    \item Powerful jets with efficiencies $\eta_\mathrm{out} > 1$ can form in extremely misaligned ($\mathcal{T}_\mathrm{init}=75^\circ$) MAD models, forcing the inner disk into alignment up to $r \gtrsim 10 r_g$ via Magneto-spin alignment \citep{Chatterjee_2023}. We have demonstrated Magneto-spin alignment at a larger initial misalignment than any previous works.

    \item SANE, or weak field misaligned simulations appear to have a maximum stable disk tilt angle of $\sim 50^\circ$ at $\sim10 r_g$, regardless of the initial disk misalignment. This peak angle is likely spin-dependent because LT precession is proportional to the spin of a black hole, however, future studies with a larger spin sampling are required to study this effect in more detail.

    \item Near the black hole, ($r \lesssim 10 r_g$), MAD and SANE models evolve very different, the jet is sufficiently efficient to quickly force the inner edge of the accretion disk into alignment if the MAD state is reached. However, at larger radii MAD and SANE models act similarly, without the powerful jet influencing the disk's evolution.

    \item A future movie campaign to observe \m87 may be able to resolve the base of the jet with the addition of intermediate baselines. It is unlikely the jet will be resolved to large enough distances from the central black hole to test if the accretion disk is misaligned or not, but it will provide an additional test of the MAD vs SANE dichotomy.

    \item We have demonstrated misaligned disk simulations can be run to extremely high initial misalignment angles with KHARMA, without any significant influence from the polar boundary on evolution of the accretion disk. Our results are consistent with comparable problems produced with \texttt{H-AMR} and \texttt{Athena++}.
    
\end{itemize}

\section*{Acknowledgements}

%The Acknowledgements section is not numbered. Here you can thank helpful
%colleagues, acknowledge funding agencies, telescopes and facilities used etc.

J.S.S would like to thank Hong-Xuan Jiang for assistance when calculating disk properties, and Paul Tiede for useful discussions regarding jet detection in EHT images.
J.S.S. acknowledges funding support from the UK STFC through grant number ST/X508329/1.
C.P. was supported in part by the Gordon and Betty Moore Foundation (Grant \#13526). It was also made possible through the support of a grant from the John Templeton Foundation (Grant \#63445).  The opinions expressed in this publication are those of the author(s) and do not necessarily reflect the views of these Foundations.
D.W.T. acknowledges funding support from the UK STFC through grant number ST/R000786/1.
H.C. was partially supported by the black hole Initiative at Harvard University, which is funded by the Gordon and Betty Moore Foundation (Grant \#8273.01). It was also made possible through the support of a grant from the John Templeton Foundation (Grant \#62286). The opinions expressed in this publication are those of the authors and do not necessarily reflect the views of these Foundations.
This work used the DiRAC Extreme Scaling service (Tursa / Tesseract [*]) at the University of Edinburgh, managed by the Edinburgh Parallel Computing Centre on behalf of the STFC DiRAC HPC Facility (www.dirac.ac.uk). The DiRAC service at Edinburgh was funded by BEIS, UKRI and STFC capital funding and STFC operations grants. DiRAC is part of the UKRI Digital Research Infrastructure.
We are grateful for use of the computing resources from the Northern Ireland High Performance Computing (NI-HPC) service funded by EPSRC (EP/T022175).
We thank Abhishek Joshi, Vedant Dhruv, C.K. Chan, and Charles Gammie for
the aligned models used here, generated under NSF grant AST 20-34306.
This research used resources of the Oak Ridge Leadership Computing Facility at the Oak Ridge National Laboratory, which is supported by the Office of Science of the U.S. Department of Energy under Contract No. DE-AC05-00OR22725. This research used resources of the Argonne Leadership Computing Facility, which is a DOE Office of Science User Facility supported under Contract DE-AC02-06CH11357. This research was done using services provided by the OSG Consortium, which is supported by the National Science Foundation awards \#2030508 and \#1836650. This research is part of the Delta research computing project, which is supported by the National Science Foundation (award OCI 2005572), and the State of Illinois. Delta is a joint effort of the University of Illinois at Urbana-Champaign and its National Center for Supercomputing Applications.
The James Clerk Maxwell Telescope (JCMT) is operated by the East Asian Observatory on behalf of the National Astronomical Observatory of Japan; the Academia Sinica Institute of Astronomy and Astrophysics; the Korea Astronomy and Space Science Institute; the National Astronomical Research Institute of Thailand; the Center for Astronomical Mega-Science (as well as the National Key R\&D Program of China with grant no. 2017YFA0402700). Additional funding support is provided by the STFC of the UK and participating universities and organizations in the UK, Canada and Ireland. The JCMT has historically been operated by the Joint Astronomy Centre on behalf of the STFC of the UK, the NRC of Canada and the Netherlands Organisation for Scientific Research.

\textit{Software}: KHARMA \citep{Prather_2021, Prather_2024}, Numpy \citep{Harris_2020}, Scipy \citep{Pauli_2020}, Pandas \citep{McKinney_2010}, Matplotlib \citep{Hunter_2007}.

%%%%%%%%%%%%%%%%%%%%%%%%%%%%%%%%%%%%%%%%%%%%%%%%%%
\section*{Data Availability}

%The inclusion of a Data Availability Statement is a requirement for articles published in MNRAS. Data Availability Statements provide a standardised format for readers to understand the availability of data underlying the research results described in the article. The statement may refer to original data generated in the course of the study or to third-party data analysed in the article. The statement should describe and provide means of access, where possible, by linking to the data or providing the required accession numbers for the relevant databases or DOIs.

The data used in this article will be made available upon request to the corresponding author.



\bibliographystyle{mnras}
\bibliography{example} % if your bibtex file is called example.bib

%\begin{thebibliography}{99}
%\end{thebibliography}

%%%%%%%%%%%%%%%%%%%%%%%%%%%%%%%%%%%%%%%%%%%%%%%%%%

%%%%%%%%%%%%%%%%% APPENDICES %%%%%%%%%%%%%%%%%%%%%

\appendix

\section{Polar Boundary Treatment}
\label{sec::appendix}

\subsection{Implementation}

Without adequate polar boundary conditions, accretion disks cannot be misaligned greater than $\sim30\degree$, unless small initial disks are used to enure no material crosses the boundary. At best, highly misaligned models will contain unphysical polar artifacts due to the build up of flux and magnetic fields at the coordinate pole, and at worst the disk will be ripped apart. Many codes have boundary conditions that do not cause a build up of flux or magnetic field at the pole. The flux over the pole is calculated with a modified implementation of the excised boundary conditions used in \texttt{H-AMR} (\citealt[Supplementary Information 2.3]{Liska_2018}). We briefly summarise our modified implementation below.

\begin{figure}
    \centering
    \includegraphics[width=\linewidth]{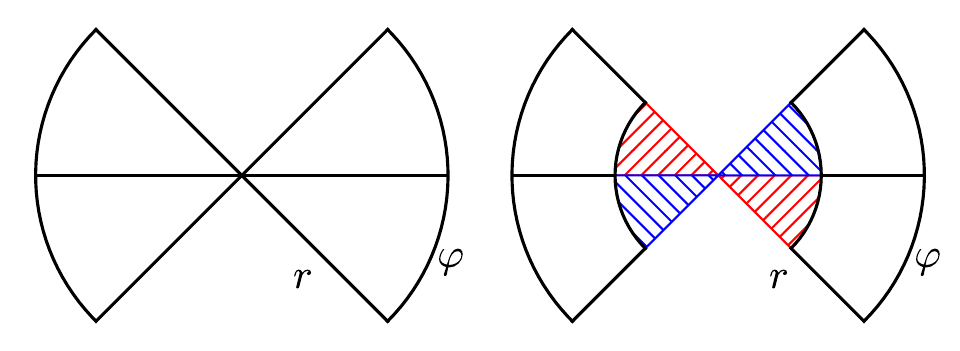}
    \caption{A schematic diagram of the half-cell treatment at the pole in our excised boundary conditions. Left: The standard cell layout at the pole in spherical coordinated systems without excised boundary conditions. Right: The cell layout when excised boundary conditions are used. Red and blue hatched sections are removed and become the excised region over which fluxes can be calculated.}
    \label{fig:excision-schematic}
\end{figure}

\begin{itemize}
    \item We ``remove'' the inner half of the first rank cell surrounding the polar singularity, leaving a half-sized face and an excised region, see Figure \ref{fig:excision-schematic} for simple a schematic. Fluxes can now be calculated across the excised zone with half-sized faces using the LLF scheme, and replace the initially calculated fluxes.
    
    \item To use the existing divergence operator, the fluxes through half-sized faces are halved to account for smaller faces. After the divergence has been calculated, the result is multiplied by two to recover the full volume of the first rank cells. Finally, the conserved variables are rescaled to reflect being defined at the centre of the full cell, instead of the half sized cell.
        
    \item Tight magnetic field loops form in relativistic jets launched from black holes, but reconnect or are blown away. However, in previous boundary condition implementations this can cause unphysical build up of the magnetic fields $\varphi$ component, $\mathrm{B}_\varphi$, at the polar boundary. Therefore, we forcibly ``reconnect'' $\mathrm{B}_\varphi$, by subtracting the average $\mathrm{B}_\varphi$ around the pole from $\mathrm{B}_\varphi$ in each cell around the pole. Reducing unphysical buildup of $\mathrm{B}_\varphi$. 
\end{itemize}

\subsection{Testing}

In Section \ref{sec::results}, our $45\degree$ and $75\degree$ models significantly intersect the polar boundary. We need to be confident that our polar boundary conditions have no significant effect on the evolution of an accretion disk. In this section, we test our boundary conditions, to ensure there is no significant effect on the accretion disks evolution. 

\begin{figure*}
    \centering
    \includegraphics[width=\linewidth]{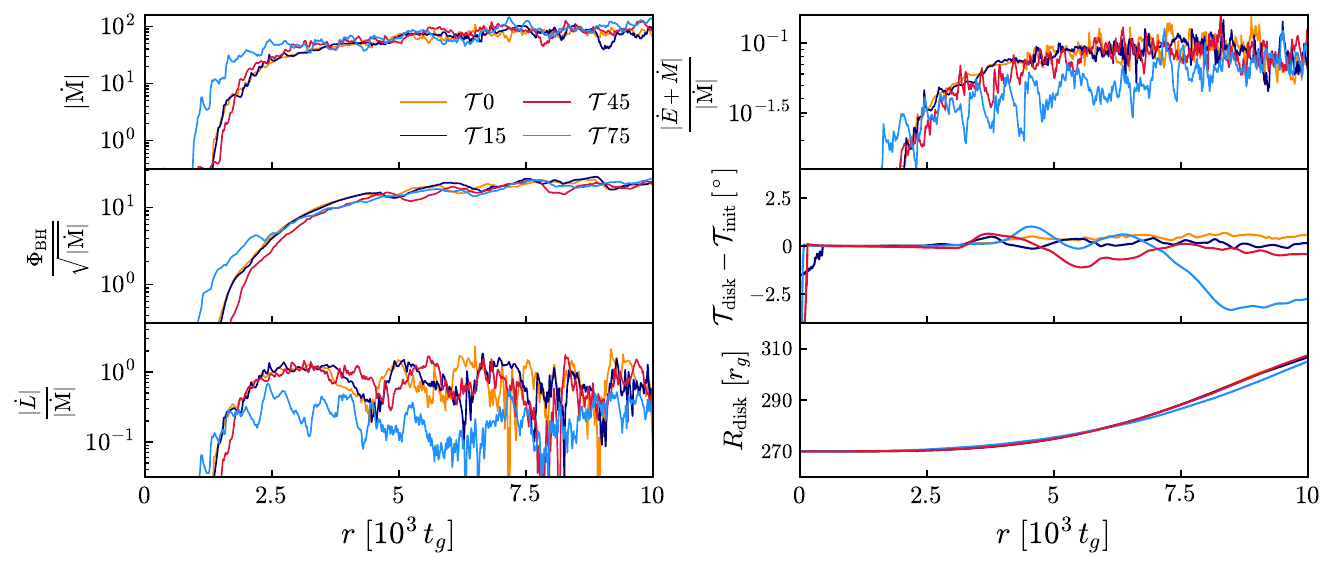}
    \caption{Temporal evolution of disk properties in our Schwarzschild simulations. From left to right, top to bottom, rest mass accretion rate, normalised angular momentum, and normalised energy with rest mass contributions removed, normalised dimensionless horizon penetrating magnetic flux, initial tilt subtracted disk tilt, normalised angular momentum, and barycentric radius. We see minimal differences beyond expected turbulent evolution of MAD models, and find our polar boundaries conditions behave well.}
    \label{fig:boundary-temporal}
\end{figure*}

For an accretion disk to align with its central BH, frame-dragging is required to apply the additional torque forces. Schwarzschild BHs, unlike Kerr BHs, have no frame dragging, and therefore we expect no alignment of the accretion disk. Both an aligned and misaligned accretion disk should evolve similarly around a Schwarzschild BH, regardless of the initial angle of misalignment. To test our boundary implementation, we have produced 4 models of an accretion disk, initially misaligned by $\mathcal{T}=0\degree, 15\degree, 45\degree, 75\degree$, evolved for $10,000t_g$, with a base resolution of $N_r \times N_\theta \times N_\varphi = 192 \times 128 \times 128$, in WKS coordinates, with $f_\mathrm{lin}=0.9$, $\lambda=0.03$, and one level of ISMR. In Figure~\ref{fig:boundary-temporal} we plot the temporal evolution of the rest mass accretion rate, dimensionless magnetic flux, angular momentum, energy outflow, normalised disk tilt, and the barycentric radius of the accretion disk. Figure~\ref{fig:boundary-temporal} shows all our models match very well within the turbulent nature of an accretion black hole in the MAD regime. In some cases, the model tilted by $75\degree$ differs slightly, however, this is due to a larger portion of the accretion disk evolving within the lower resolution polar region. If the difference was due to the boundary conditions, the $45\degree$ model would also differ as it also significantly crosses the polar boundary. The lower resolution polar region may have some influence in our production simulations, however, the increased resolution will reduce the effects seen in Figure~\ref{fig:boundary-temporal}. Additionally, the difference at large misalignments is minimal, only $\sim2.5\degree$, which does not influence the conclusions of this paper. We see a similar impact from polar boundary conditions as the \texttt{H-AMR} implementation (\citealt[Appendix 6.2]{Chatterjee_2023}). 

\begin{figure*}
    \centering
    \includegraphics[width=\linewidth]{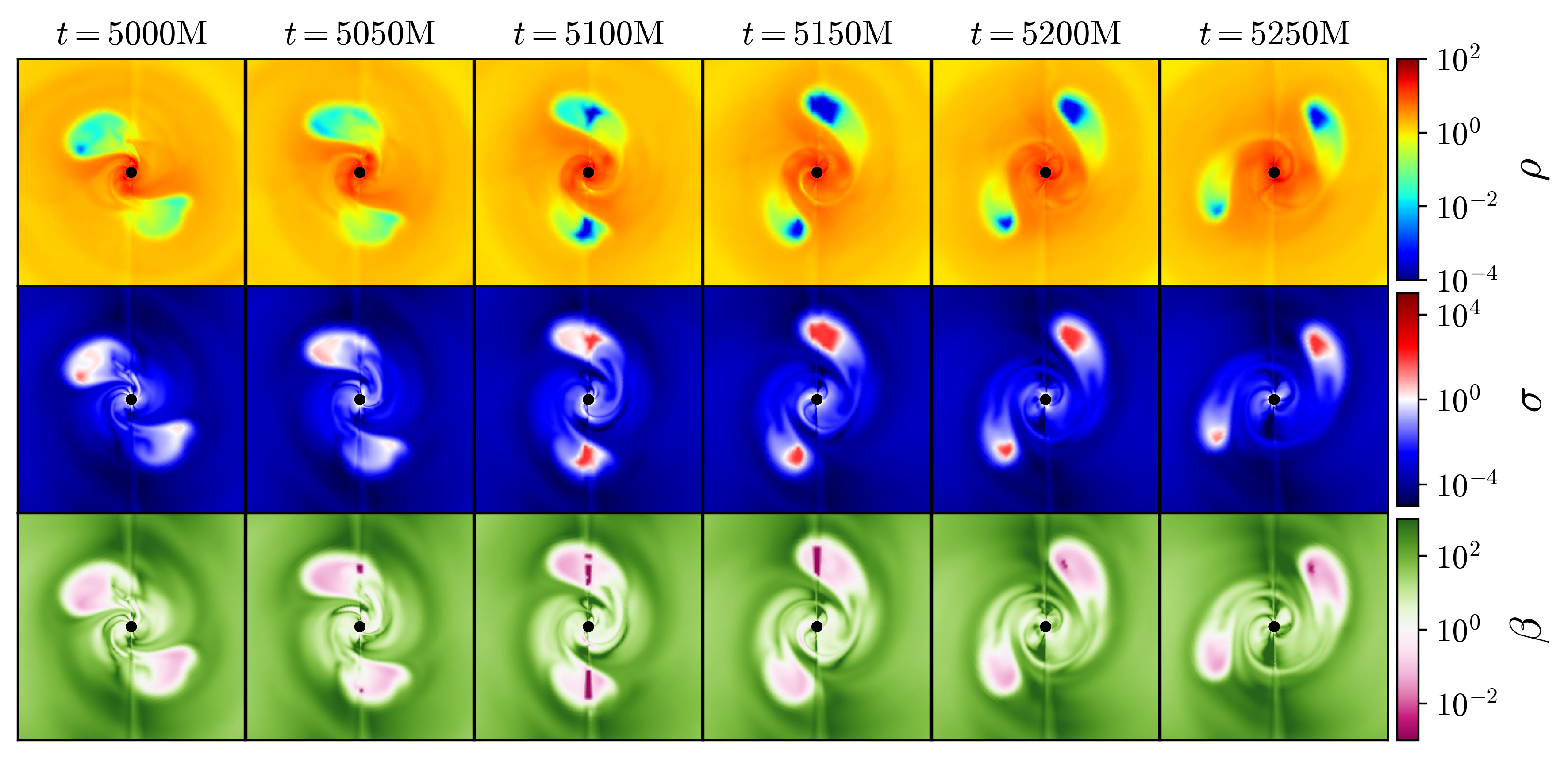}
    \caption{A midplane slice of density $\rho$, magnetisation $\sigma$, and plasma $\beta$ as a flux tubes crosses the polar boundary. This model is a $90\degree$ misaligned MAD accretion disk with a base resolution of $N_r \times N_\theta \times N_\varphi = 128 \times 64 \times 64$, around a Schwarzschild BH, produced in eKS coordinates, with 0 levels of ISMR. Each panel shows the inner $60 \, r_\mathrm{g} \times 60 \, r_\mathrm{g}$ of the grid.}
    \label{fig:flux-tube}
\end{figure*}

Flux tubes are the low density, highly magnetised evacuated regions within the accretion disk produced during flux eruption events in MAD models. The sharp gradient in density, magnetisation and plasma $\beta$ present a potential challenge as a flux tube crosses the polar boundary in highly tilted models. In Figure \ref{fig:flux-tube} we plot the density, magnetisation and plasma $\beta$ at $50  t_g$ increments during a flux eruption event. 

The reduction in $\rho$ and $\sigma$ is due to the polar wake parting as it crosses the polar boundary. As the flux tube crosses the boundary, some of the material is slightly deflected of the midplane, which causes the drop in $\rho$, and increase in $\sigma$. The low plasma $\beta$ is a side effect of the $\mathrm{B}_\varphi$ reconnection, magnetic field energy is turned into thermal energy at the next primitive variable reconstruction. However, it is clear from Figure~\ref{fig:flux-tube}, neither of these effects are dynamically important.

The current boundary condition implementation would not be recommended for a detailed investigation of flux tubes in a highly misaligned accretion disk, however, it is clear the general structure and evolution of the accretion disk, which we study in this paper, is uneffected by the current implementation. We are therefore confident the polar boundary implementation is sufficient to achieve the goals of this paper.

\section{Effect of Varying the Initial Conditions}
\label{sec::AppendixB}

To confirm our hypothesis of the adiabatic index being responsible for \texttt{MAD75} aligning, despite the very large initial disk tilt, we ran an additional simulation with $\gamma=4/3$ and all other parameters fixed. We also ran a further simulation equivalent to \texttt{MAD75} but with a smaller initial disk size. In this simulation we placed the inner edge of the accretion disk at $10r_g$ and the pressure maximum at $20 r_g$. Each of these simulations were evolved for $10,000 t_g$.

\begin{figure}
    \centering
    \includegraphics[width=\linewidth]{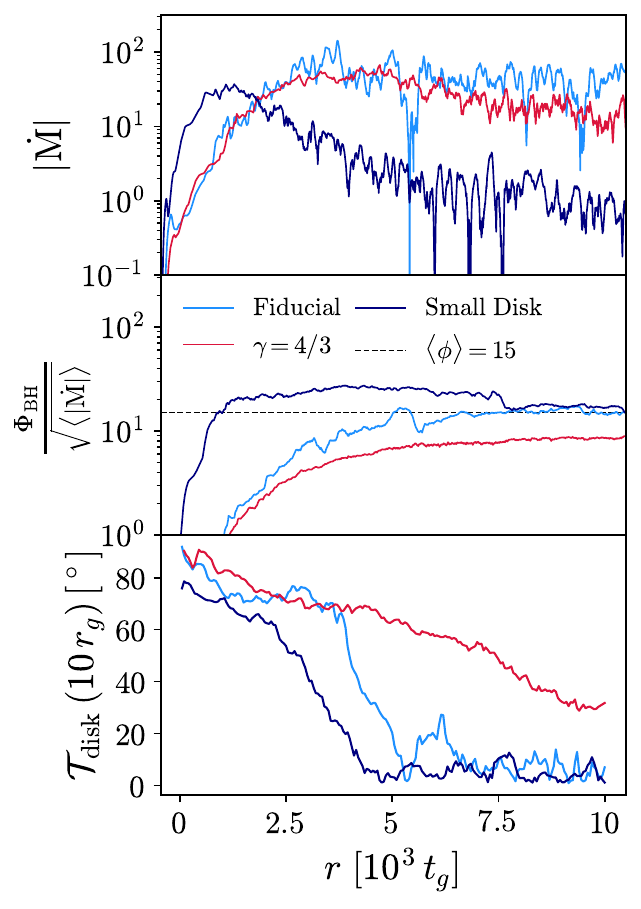}
    \caption{Comparisons of the mass accretion rate (top row), dimensionless magnetic flux crossing the event horizon (middle row), and angle of the accretion disk measured at $10 r_g$ (bottom row) for our fiducial (light blue), small disk (dark blue), and $\gamma=4/3$ (red) \texttt{MAD75} simulations. The size of accretion disk has minimal impact on the evolution of the accretion disk, whereas, a smaller value of the adiabatic index may not reach the MAD state at large initial disk tilt, leading the large evolutionary differences.}
    \label{fig:diff-init}
\end{figure}

Figure~\ref{fig:diff-init} shows the mass accretion rate, dimensionless magnetic flux crossing the event horizon, and the disk tilt measured at $10 r_g$ from the event horizon in our fiducial, small disk, and $\gamma=4/3$ \texttt{MAD75} simulations. We find the small disk model acts similarly to the fiducial model, reaching the MAD state and forcing the inner disk into alignment with the black hole within $\sim5,000 t_g$. The smaller disk reaches the MAD state earlier in the simulation, resulting in near disk alignment occurring earlier in the run compared to the fiducial simulation. The smaller gas reservoir in the small disk model leads to decreasing mass accretion rate faster than the fiducial simulation. The size of the accretion disk does not strongly influence the evolution of misaligned accretion disks.

Our simulation with $\gamma=4/3$ does not reach the MAD state, the disk becomes magnetically saturated at $\langle \phi \rangle \sim 5$, below the MAD threshold. Additionally, the inner disk in this model does not align with the black hole within $10,000 t_g$, unlike our $\gamma=5/3$ simulations. We observe partial alignment of the inner disk (Figure~\ref{fig:diff-init} bottom panel), similar to the early evolution of \texttt{T75} (see Figure 3 of \citealt{Chatterjee_2023}). We expect, if we evolved this simulation for a longer duration, we would find the same late-time behaviour as \texttt{T75}, a misaligned, weakly magnetised accretion disk. 

\begin{figure}
    \centering
    \includegraphics[width=\linewidth]{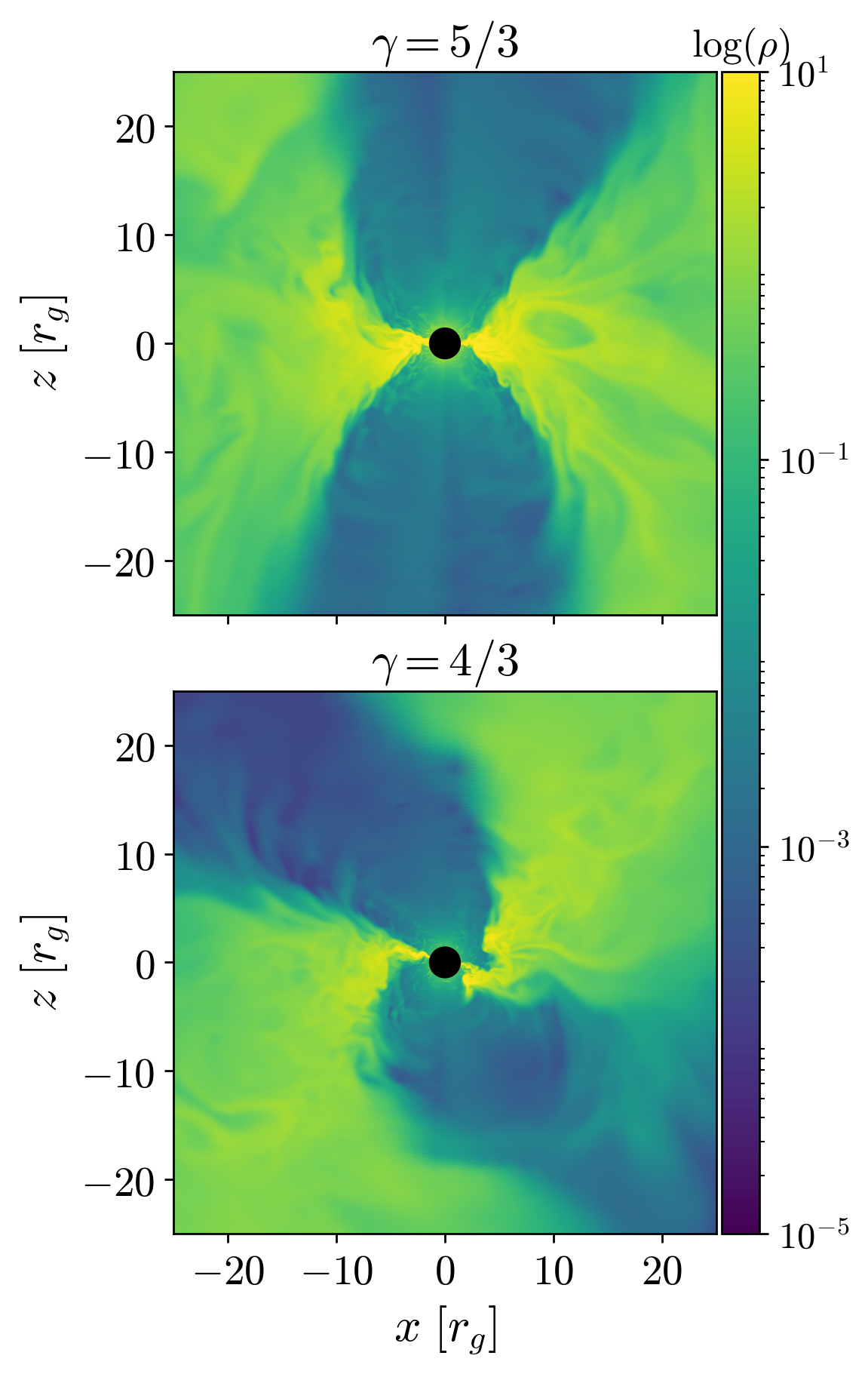}
    \caption{Snapshot images at $10,000 t_g$ for our fiducial \texttt{MAD75} simulation with $\gamma=5/3$ (top panel) and our $\gamma=4/3$ (bottom panel) simulation, the adiabatic index is the only different input parameter between simulations. Structural differences are clear between each simulation, larger values of the adiabatic index can fully align the inner accretion disk, where smaller values cannot.}
    \label{fig:diff-init-snapshot}
\end{figure}

Figure~\ref{fig:diff-init-snapshot} shows clear structural differences between extremely misaligned models evolved for different adiabatic indices, with all other parameters fixed. The choice of adiabatic index for misaligned simulations should be carefully considered, we have shown drastic differences in misaligned simulations when changing the adiabatic index. Future misaligned simulations should be as realistic when selecting the adiabatic index, \citep{Chael_2025} and \cite{Gammie_2025} suggest this should be $\gamma=5/3$.

% comment about the importance of adiabatic index in misaligned models, add comment to discussion as well.

%%%%%%%%%%%%%%%%%%%%%%%%%%%%%%%%%%%%%%%%%%%%%%%%%%

% Don't change these lines
\bsp	% typesetting comment
\label{lastpage}
\end{document}